\documentstyle[11pt]{article}
\begin{document}
\thispagestyle{empty}
\baselineskip 18pt
\rightline{UOSTP-99-008}
\rightline{{\tt hep-th/9910135}}

\

\def\tr{{\rm tr}\,} \newcommand{\beq}{\begin{equation}}
\newcommand{\eeq}{\end{equation}} \newcommand{\beqn}{\begin{eqnarray}}
\newcommand{\eeqn}{\end{eqnarray}} \newcommand{\bde}{{\bf e}}
\newcommand{\balpha}{{\mbox{\boldmath $\alpha$}}}
\newcommand{\bsalpha}{{\mbox{\boldmath $\scriptstyle\alpha$}}}
\newcommand{\betabf}{{\mbox{\boldmath $\beta$}}}
\newcommand{\bgamma}{{\mbox{\boldmath $\gamma$}}}
\newcommand{\bbeta}{{\mbox{\boldmath $\scriptstyle\beta$}}}
\newcommand{\lambdabf}{{\mbox{\boldmath $\lambda$}}}
\newcommand{\bphi}{{\mbox{\boldmath $\phi$}}}
\newcommand{\bslambda}{{\mbox{\boldmath $\scriptstyle\lambda$}}}
\newcommand{\ggg}{{\boldmath \gamma}} \newcommand{\ddd}{{\boldmath
\delta}} \newcommand{\mmm}{{\boldmath \mu}}
\newcommand{\nnn}{{\boldmath \nu}}
\newcommand{\diag}{{\rm diag}}
\newcommand{\bra}[1]{\langle {#1}|}
\newcommand{\ket}[1]{|{#1}\rangle}
\newcommand{\sn}{{\rm sn}}
\newcommand{\cn}{{\rm cn}}
\newcommand{\dn}{{\rm dn}}
\newcommand{\tA}{{\tilde{A}}}
\newcommand{\tphi}{{\tilde\phi}}
\newcommand{\bpartial}{{\bar\partial}}
\newcommand{\br}{{{\bf r}}}

\

\vskip 0cm
\centerline{\large\bf Deformed Nahm Equation and a 
Noncommutative BPS Monopole}

\vskip .2cm

\vskip 1.2cm
\centerline{\large\it
Dongsu Bak
% $^a$
\footnote{Electronic Mail: dsbak@mach.uos.ac.kr}
%and Kimyeong Lee $^{bc}$\footnote{Electronic Mail:
%klee@kias.re.kr} 
}
\vskip 5mm
\centerline{ \it 
%$^a$ 
Physics Department,
University of Seoul, Seoul 130-743, Korea}
\vskip 3mm

\vskip 1.2cm
\begin{quote}
{\baselineskip 16pt
A deformed Nahm equation for the BPS equation
in the noncommutative N=4 supersymmetric U(2) 
Yang-Mills 
theory is obtained. Using this, we constructed 
explicitly a monopole solution of 
the noncommutative BPS equation to the linear 
order of the 
noncommutativity scale.
We found that 
the leading order correction to the 
ordinary SU(2)
monopole  lies solely in the overall U(1)
sector and that the overall  U(1) magnetic
field has an expected long range component of 
magnetic dipole moment.   
\\
(14.80.Hv,11.15.-q)
}
\end{quote}

%\pacs{14.80.Hv,11.27.+d,14.40.-n}

\newpage

\baselineskip 12pt

As shown recently, 
quantum field theories in noncommutative spacetime
naturally arise as a decoupling limit of the 
worldvolume dynamics of D-branes in a constant NS-NS 
two-form background\cite{seiberg}. 
But detailed dynamical effects due to the 
noncommutative geometry are still partially 
understood\cite{seiberg,connes,douglas,itzhaki}. 

In this note,
we shall exploit 
the N=4 supersymmetric gauge theories on noncommutative 
${\bf R}^4$ that is the worldvolume theory of D3-branes
in the NS-NS two-form background. This theory
was recently investigated to study the nature of 
monopoles and dyons
in the noncommutative space\cite{hashimoto}. 
The energy and charge of 
 monopoles satisfying the noncommutative BPS equations were 
identified and shown to agree those of ordinary 
monopoles. 
Below, we shall concentrate on the construction of 
a self-dual monopole solution of the theory by generalizing
ADHMN methods\cite{nahm1,hitchin,bowman,corrigan} to the noncommutative case.
%that satisfies the noncommutative 
%BPS equation.      

We begin with the noncommutative 
 N=4  supersymmetric Yang-Mills 
theory.
We shall restrict our discussion to the case of  $U(2)$ gauge group.
Among the six Higgs fields, only a Higgs field $\phi$ plays a role
in the following discussions of  a  monopole. 
The  bosonic part of the action is given  by
\begin{equation}
S= -\frac{1}{4g^2_{\rm YM}} \int d^4x\; {\rm tr} \Bigl( F_{\mu\nu} *
 F^{\mu\nu}-2 D_\mu\phi * D^\mu\phi \Bigr),
\label{lag}
\end{equation}
where the $*$-product is defined by
\begin{equation}
a(x)* b(x)\equiv \Bigl(e^{{i\over 2}\theta^{\mu\nu}\partial_\mu 
\partial'_\nu} a(x) b(x')\Bigr){\Big\vert}_{x=x'}
\label{star}
\end{equation}
that respects the associativity of the product. We shall assume in the following 
that  $\theta_{0i}=-\theta_{0i}=0$. Then without loss of generality, one may
take the only nonvanishing components to be 
$\theta_{12}=-\theta_{21}\equiv\theta$.
$F_{\mu\nu}$ and $D_\mu\phi$
are defined respectively by
\begin{eqnarray}
&&F_{\mu\nu}\equiv \partial_\mu A_\nu-\partial_\nu A_\mu
+i(A_\mu * A_\nu-A_\nu * A_\mu)\nonumber\\ 
&&D_{\mu}\phi\equiv \partial_\mu \phi 
+i(A_\mu * \phi-\phi *A_\mu)
\label{fieldstrength}
\end{eqnarray}
The four vector potential and $\phi$ 
belong to $U(2)$ Lie algebra given by
$T_{4}={1\over {2}}I_{2\times 2}$ and 
$(T_1,T_2,T_3)={1\over {2}}(\sigma_1,\sigma_2,\sigma_3)$ normalized 
by  $\tr T_m T_n={1\over 2}\delta_{mn}$. The vacuum expectation value
of the Higgs field $\langle \phi\rangle$ is taken to be $T_3 U$ in 
the asymptotic region. % in an appropriate gauge.

As shown in Ref.~\cite{hashimoto}, the energy functional 
\begin{eqnarray}
M\!=\! {1\over 2 g^2_{\rm YM}}\!\!
\int\! d^3x \,\tr \Bigl(E_i * E_i \!+\! D_0\phi * 
D_0\phi\!+\! B_i * B_i\!+\! D_i\phi * D_i\phi\Bigr) \ge {1\over g^2_{YM}}
\int_{r=\infty}\!\! dS_k \tr B_k * \phi,   
\label{energy}
\end{eqnarray}
is bounded as in the case of the ordinary supersymmetric Yang-Mills theory.
 The saturation of the bound occurs when
the  BPS equation 
\begin{eqnarray}
B_i=D_i\phi
\label{bps}
\end{eqnarray}
is satisfied. The mass for the solution  is 
\begin{eqnarray}
M= {2\pi Q_M \over g^2_{\rm YM}} U
\label{mass}
\end{eqnarray}
where we define the magnetic charge $Q_M$  by
\begin{eqnarray}
Q_M={1\over 2\pi U} \int_{r=\infty} dS_k \tr B_k \phi.
\label{charge}
\end{eqnarray} 
As argued in Ref.~\cite{hashimoto}, the charge is
 to be quantized at integer values even 
in the noncommutative case. This is because the fields in the 
asymptotic region are slowly varying and, hence, the standard
argument of the topological quantization of the magnetic charge holds
in the noncommutative theory. 
The main aim of this note is to investigate the detailed form of 
one self-dual monopole solution in the noncommutative case. 

Before proceeding we 
show 
first the fact that a noncommutative 
monopole solution inevitably involves nonvanishing
overall $U(1)$  parts.  To prove this, let us note
the solution for the monopole with $\theta=0$ is given
 by\footnote{In the following,
we shall set $g_{\rm YM}=1$ and $U=1$.}
\begin{eqnarray}
&&\tphi\ ={\bf r} \cdot\sigma {1\over 2r}\left(\coth r-{1\over r}\right)\\
\label{higgssol}
&&{\tA}_i=\epsilon_{ijk}x^j\sigma^k {1\over 2 r^2}
\left({r\over \sinh r} -1\right)\,.
\label{gaugessol}
\end{eqnarray} 
We then expand the solution of the noncommutative BPS equation with respect to
 $\theta$ by
\begin{eqnarray}
&&\phi=\tphi +\theta\phi_{(1)}+\theta^2\phi_{(2)}\cdots\\
\label{higgs}
&&A=\tA+\theta A_{(1)}+\theta^2 A_{(2)}\cdots\,.
\label{gauge}
\end{eqnarray} 
The leading corrections to $D_i\phi$ and $B_i$ are, respectively,
\begin{eqnarray}
&& (B_i)_{(1)}=\epsilon_{ijk}\left(i(\partial\tA_j\bpartial\tA_k-
\bpartial\tA_j\partial\tA_k)+\partial_j (A_{(1)})_k+i[\tA_j,(A_{(1)})_k]\right)
\label{gauge1}\\
&&(D_i\phi)_{(1)}=i(\partial\tA_i \bpartial\tphi 
\!-\!\bpartial\tA_i \partial\tphi 
\!-\!\partial\tphi \bpartial\tA_i \!+\!\bpartial\tphi \partial\tA_i)
\!+\!\partial_i \phi_{(1)}
\!+\!i[ (A_{(1)})_i,\tphi]\!+\!i[ \tA_i,\phi_{(1)}]
\,,
\label{higgs1}
\end{eqnarray} 
where $\partial$ and $\bpartial$ denote derivative with respect to
$x+iy$ and $x-iy$.
The terms in the first parentheses follow from the evaluation of the 
$*$-product of the ordinary monopole solution and can be 
explicitly computed. They are 
\begin{eqnarray}
&& i\epsilon_{ijk}\left((\partial\tA_j\bpartial\tA_k-
\bpartial\tA_j\partial\tA_k)\right)={1\over 2r}
\left(-{(r^2 w^2)'} \delta_{3i}
+{(w^2)'}x^3x^i\right)I_{2\times 2}
\label{gauge11}\\
&&i(\partial\tA_i \bpartial\tphi -\bpartial\tA_i \partial\tphi 
-\partial\tphi \bpartial\tA_i -\bpartial\tphi \partial\tA_i)
={1\over r}\left(-{(r^2 w h)'} \delta_{3i}
+{(wh)'}x^3x^i\right)I_{2\times 2}
\,,
\label{higgs11}
\end{eqnarray} 
where
%\begin{eqnarray}
%&& 
$w={1\over 2 r^2}\left({r\over \sinh r} -1\right)$
%\label{gaugeaa}\\
%&&
and $h={1\over 2r}\left(\coth r-{1\over r}\right)$.
%\,.
%\label{higgsaa}
%\end{eqnarray} 
We note that these have overall $U(1)$ components only 
and the two are different from each other. Hence to satisfy the 
Bogomol'nyi equation, there should be overall $U(1)$ contributions in 
$\phi_{(1)}$ or $A_{(1)}$ to cancel the difference because the 
$SU(2)$ parts of $\phi_{(1)}$ and $A_{(1)}$ do not produce $U(1)$
components of the field strengths in this leading order. 

We now establish the Nahm's formalism in the noncommutative case
to solve the BPS equation. The basic idea is simply to replace all 
the ordinary product of the standard derivation by $*$-product.
This fact was briefly argued already in Ref.~\cite{hashimoto}, but the
deformed Nahm equation below was not found.
To obtain the self-dual  monopole solution, Nahm adapted the 
ADHM construction\cite{ADHM} of instanton\footnote{See also 
Ref.\cite{nekrasov} for the ADHM construction of noncommutative 
instanton solutions.} 
to solve the self-dual Yang-Mills
equation in Euclidean four space such that the gauge fields are
translationally invariant in the $x_4$-direction. Here we shall
begin with reviewing ADHMN construction\cite{nahm1,hitchin,bowman,corrigan}
to follow the similar line of logic. 
The ADHMN construction of $k$ monopole
starts with a matrix operator,
\begin{eqnarray}
\Delta^\dagger=
-{d\over d\tau} I_{k\times k}\otimes I_{2\times 2} +
 I_{k\times k}\otimes\sigma_i x_i  +T_i \otimes\sigma_i  
\,,
\label{projection}
\end{eqnarray}
where $T_i$'s are $k\times k$ matrices.
The following two conditions are required on $\Delta$;
\begin{eqnarray}
[\Delta^\dagger \Delta, I_{k\times k}\otimes \sigma_i]=0
\,,\ \ \ \ 
%\label{condition1}\\
\Delta^\dagger \Delta: \ {\rm invertible}\,.
\label{condition1}
\end{eqnarray}
We then solve  the equation
\begin{eqnarray}
\Delta^\dagger V=0
\,,
\label{deltaequation}
\end{eqnarray}
with a normalization condition 
\begin{eqnarray}
\int^{{1\over 2}}_{-{1\over 2}}\!\! d\tau\,\,   V^\dagger V =I_{2k\times 2k}
\,,
\label{nomalization}
\end{eqnarray}
where $V$ is a $2k\times 2k$ matrix.
The gauge fields are given by
\begin{eqnarray}
\tphi=\int^{{1\over 2}}_{-{1\over 2}}\!\! d\tau\,\, \tau  V^\dagger V 
\,,\ \ \ \ 
%\label{nahmhiggs}
%\\
\tA_i=-i\int^{{1\over 2}}_{-{1\over 2}}\!\! d\tau\,\,  
V^\dagger \partial_i V \,.
\label{nahmpotential}
\end{eqnarray}
One may  directly verify that  
\begin{eqnarray}
F_{mn}=2\bar\eta^i_{mn}\int^{{1\over 2}}_{-{1\over 2}}\!\! d\tau 
\!\int^{{1\over 2}}_{-{1\over 2}}\!\! d\tau' 
\,\,V^\dagger (\tau)I_{k\times k}\otimes \sigma_i (\Delta^\dagger 
\Delta)^{-1}
(\tau,\tau') V(\tau')
\label{selfduality}
\end{eqnarray}
where the indices run from 1 to 4, we  identify $A_4$ with $\phi$,
i.e. $F_{i4}=D_i\phi$,
and $\bar\eta^i_{mn}= \epsilon_{imn4}-\delta_{im}\delta_{4n}+
\delta_{in}\delta_{4m}$ is the self-dual 't Hooft tensor.

In the noncommutative case, the derivation goes through once all
the product operations  are replaced by $*$-product operations. 
Namely, the matrix operator (\ref{projection})
remains the same while the conditions become
\begin{eqnarray}
[\Delta^\dagger * \Delta, I_{k\times k}\otimes \sigma_i]=0\,,
\ \ \ \ 
%\label{condition11}\\
\Delta^\dagger * \Delta: \ {\rm invertible}\,.
\label{condition22}
\end{eqnarray}
We have to solve 
\begin{eqnarray}
\Delta^\dagger * V=0
\,,
\label{deltaequation11}
\end{eqnarray}
with a normalization condition 
\begin{eqnarray}
\int^{{1\over 2}}_{-{1\over 2}}\!\! d\tau\,\,   V^\dagger * V =I_{2k\times 2k}
\,.
\label{nomalization11}
\end{eqnarray}
The gauge fields are now given by
\begin{eqnarray}
\phi=\int^{{1\over 2}}_{-{1\over 2}}\!\! d\tau\,\, \tau  V^\dagger * V 
\,, \ \ \ \ \ 
%\label{nahmhiggs11}
%\\
A_i=-i\int^{{1\over 2}}_{-{1\over 2}}\!\! d\tau\,\,  V^\dagger * \partial_i V 
\label{nahmpotential11}
\end{eqnarray}
These potentials then satisfy the noncommutative BPS equation.
Namely, the field strength is the one obtained from 
(\ref{selfduality})  by replacing all the ordinary 
product operations with $*$-product operations, so the expression is
manifestly self-dual.

Let us now work out the conditions in (\ref{condition1}). 
%and
%(\ref{condition2}). 
The operator $\Delta^\dagger *\Delta$
reads
\begin{eqnarray}
&&\Delta^\dagger *\Delta=-{d^2\over d\tau^2}I_{k\times k}\otimes I_{2\times 2}  
+(T_i +x^iI_{k\times k})
(T_i^\dagger +x^iI_{k\times k})\otimes I_{2\times 2}\nonumber\\
&&\ \ \ \ +(T_i-T_i^\dagger)\otimes \sigma_i{d\over d\tau}
+(-{d\over d\tau} T^\dagger_i +i\epsilon_{ijk}T_j T^\dagger_k -
\theta \delta_{3i}I_{k\times k})\otimes \sigma_i
\,.
\label{deltasquare}
\end{eqnarray}
The two conditions demand that $T_i$'s are Hermitian matrices 
and that
\begin{eqnarray}
{d\over d\tau} T_i=i\epsilon_{ijk}T_j T_k 
-\theta \delta_{3i}I_{k\times k}
\,.
\label{deformed}
\end{eqnarray}
This equation is nothing but the Nahm equation for the ordinary Yang-Mills
theory if $\theta$ is set to zero. This deformed Nahm equation
can be reduced to the ordinary Nahm equation 
\begin{eqnarray}
{d\over d\tau} \tilde{T}_i=i\epsilon_{ijk}\tilde{T}_j \tilde{T}_k 
\,.
\label{nahmeq}
\end{eqnarray}
 introducing $\tilde{T}_i$
by $T_i=\tilde{T}_i-\theta \tau\delta_{i3}I_{k\times k}$.
The boundary conditions at $\tau=\pm1/2$ for the Nahm 
equation (\ref{nahmeq}) 
are not to be changed from those of ordinary monopoles
 even for 
the noncommutative case because
they correspond to the long range effects and
the noncommutativity plays no role. Hence the boundary 
conditions are that $\tilde{T}_i$'s have simple poles 
at the boundaries and the residues form an irreducible 
representation of $SU(2)$. The ordinary Nahm equation is
naturally interpreted as describing supersymmetric 
ground states for the worldvolume theory of suspended D-strings
between D3-branes\cite{diacunescu}. In this context, the modification 
$-\theta \tau\delta_{i3}I_{k\times k}$ implies that the D-strings 
are slanted with a slope $\theta$ where $\tau$ is identified
with a spatial coordinate of the D-string worldvolume.
The emerging picture is quite consistent with
the direct analysis of a D1-brane in the NS-NS two-form 
background where the same slope 
appears\cite{hashimoto}. Although interesting, we won't exploit
the relationship further here.

We now turn to the problem of the construction of actual 
solutions of the BPS equation. We  try the case of $k=1$, i.e.
one monopole. The deformed Nahm equation is solved trivially
by $T_i=-\tau \delta_{3i}-c_i$, where $c_i$'s
are constants related to the monopole position. We shall
set $c_i$ to zero by invoking the translational invariance of the
noncommutative Yang-Mills theory. 

The equation (\ref{deltaequation11}) is explicitly
\begin{eqnarray}
-{d\over d\tau}   V +
 \sigma_i x_i * V  - \theta\tau \sigma_3 V=
-{d\over d\tau}   V +
 \sigma_i x_i V +\theta \hat{O}V=0
\,,
\label{oneprojection}
\end{eqnarray}
where we define $\hat{O}$ by
\begin{eqnarray}
\hat{O}=
{i\theta\over 2}(\sigma_1 \partial_2-\sigma_2 \partial_1) - 
\theta\tau \sigma_3 
\,. 
\label{operator}
\end{eqnarray}
This equation is equivalent to 3-dimensional Dirac equation with time 
dependent mass in the background 
magnetic field $\theta$\footnote{The roles  of 
 magnetic field in open string theory
and noncommutative geometry
were explored in Ref.\cite{susskind}. }. 
The solution with $\theta=0$ is simply
\begin{eqnarray}
\tilde{V} (\tau,\br)=e^{\sigma\cdot \br\tau} K(r)
\,, 
\label{solutionzero}
\end{eqnarray}
where $K(r)=\left(r\over \sinh r\right)^{1/2}$ is 
determined by
the normalization condition, (\ref{nomalization}).
One may  solve the equation perturbatively 
by the Dyson series; the solution reads
\begin{eqnarray}
&&V=\tilde{V} +
\theta e^{\sigma\cdot \br\tau} (
\int^\tau_0 ds_1 \hat{O}_I (s_1)K(r)+W_{1}(\br))\nonumber\\
&&\ \ +\theta^2 e^{\sigma\cdot \br\tau}\left(
\int^\tau_0 ds_1 \hat{O}_I (s_1)
\Bigl(\int^{s_1}_0 ds_2 \hat{O}_I (s_2) K(r)+W_{1}(\br)\Bigr) +W_2(\br)\right)
+\cdots
\,, 
\label{solutionzeropert}
\end{eqnarray}
where $\hat{O}_I (\tau)=e^{\sigma\cdot \br\tau} 
\hat{O} (\tau) e^{-\sigma\cdot \br\tau}$.
All the integration constants $W_{n}(\br)$ are determined
by the normalization condition, (\ref{nomalization11}). 

Explicit evaluation of $W_{1}$ by imposing 
the normalization condition leads to
\begin{eqnarray}
W_{1}(\br)
= {K(r)\over 4r^2} \left( 
\Bigl({r\varphi }-(S-1)\Bigr)\sigma_3
+{x_3\over r}\Bigl({r^2\over 2}-{r\varphi}+(S-1)\Bigr)\sigma\cdot \hat{r}
\right)
\,, 
\label{w11}
\end{eqnarray}
where $\varphi\equiv \coth r -{1\over r}$ and $S\equiv {r\over \sinh{r}}$.
The solution, $V_{(1)}$ reads explicitly
\begin{eqnarray}
&&{V_{(1)}\over K}
= -{x_3\over 4r^2}\left( 2\tau \cosh 
(\tau r)-{\sinh(\tau r)\over r}(r\varphi+2) 
\right)
%I_{2\times 2} 
-\sinh(\tau r) 
(2\tau +\varphi \sigma\cdot\hat{r}){\sigma_3\over 4r}
\nonumber\\
&&\ \ \ -{ \tau^2x_3 \over 2r}  \sigma\cdot\hat{r}
e^{\sigma\cdot \br\tau}
+{1\over 2r^3} \Bigl(r\tau e^{-\sigma\cdot \br\tau}
-\sinh(\tau r)\Bigr)\left(\sigma\cdot \br \, \sigma_3-x_3\right) 
\nonumber\\
&&\ \ \ +{e^{\sigma\cdot\br \tau}\over 4r^2} \left( 
\Bigl({r\varphi }-(S-1)\Bigr)\sigma_3
+{x_3\over r}\Bigl({r^2\over 2}-{r\varphi}+(S-1)\Bigr)\sigma\cdot \hat{r}
\right)
\,. 
\label{v1}
\end{eqnarray}
It is then straightforward to evaluate $\phi_{(1)}$ and 
$A_{(1)}$ using (\ref{nahmpotential11}); they are
\begin{eqnarray}
&&\phi_{(1)}
= 0\,,
\label{phi1}\\
&& (A_{(1)})_i={(S-1)\over 8 r^4} \Bigl( 2r\varphi -
(S-1)\Bigr)\epsilon_{3ij}x_j I_{2\times2}
\,. 
\label{a1}
\end{eqnarray}
In this order, there is no  correction
 to the Higgs field, 
which provides the higher dimensional geometric
picture of monopole as a D-string suspended between two D3-branes. 
Finally, using (\ref{gauge1})-(\ref{higgs1}),  the leading corrections to the
magnetic field and $D\phi$ are found to be
\begin{eqnarray}
&&B_{(1)}=(D\phi)_{(1)}= \left(-{1\over r} (r^2 G)'\hat{e}_3
+ G' x_3 \hat{r}\right) I_{2\times 2}
\,,
\label{b1}
\end{eqnarray}
where $G={(S-1)\varphi\over 4r^3}$.
A few comments are in order.
First, the self duality to this order is manifest in the above as it should be.
As proved earlier, there are nonvanishing overall $U(1)$  components in
the gauge connection. Furthermore, all the leading order corrections lie
in the overall  $U(1)$ sector.
The solution in (\ref{phi1})-(\ref{a1}) is rather unique up to 
zero-mode and gauge fluctuations that may be 
shown to agree with those of ordinary SU(2) monopole 
to $O(\theta)$. Note that the translational zero modes of a monopole 
are already fixed by
setting $c_i$ to zero. 
The gauge fields, Higgs field and field strengths are nonsingular everywhere 
to this order. Especially, the magnetic field at the origin behaves
\begin{eqnarray}
\delta B =\theta \Bigl({1\over 36}\hat{e}_3 +O(r^2)\Bigr)I_{2\times 2}
\,.
\label{b1origin}
\end{eqnarray}
We see here constant U(1) magnetic field is induced at the origin.
In the asymptotic region, the correction 
behaves as\footnote{The dipole term in this expression was first
 obtained in Ref.~\cite{hata}.}  
\begin{eqnarray}
&&\delta B = \theta {-\hat{e}_3 + 3 
(\hat{e}_3\!\!\cdot \hat{r}) \hat{r}\over 4r^3}I_{2\times 2}
+\theta {\hat{e}_3- 2 
(\hat{e}_3\!\!\cdot \hat{r}) \hat{r}\over 2r^4}I_{2\times 2}
+{\rm exponential\ corrections}
\,.
\label{b1infinity}
\end{eqnarray}
This long range field does not contribute to the magnetic 
charge in (\ref{charge})
and, thus, we explicitly see that the magnetic charge remains 
at integer values.
The leading overall U(1) correction is the expected 
magnetic dipole contribution  of the form,
 ${-{\bf p}+ 3 ({\bf p}\cdot \hat{r})\hat{r}\over r^3}$, with ${\bf p}=
{\theta\over 4}\hat{e}_3 I_{2\times 2}$; As discussed below (\ref{nahmeq}),
$\pm{\theta\over 2}{\hat{e}_3}$ are the end-point displacements of the 
D-string from the  positions of $\theta=0$, and the U(1) magnetic charges with 
a proper normalization are
$\pm{1\over2}$ on the branes, so ${\bf p}$ agrees with the charges 
multiplied by the displacements. 
%but its presence reflects the fact 
%that there is significant change 
%of magnetic charge distribution around 
%the core.
The subleading long range term is not in the form of 
quadrupole moment 
and its physical origin is not clear.

There are many directions to go further.  In the large $\theta$ limit,
the structure of the equation (\ref{oneprojection}) seems simplified
and this may be helpful in understanding the nature of large $\theta$
limit by finding details.
As said earlier, one may equivalently
describe the noncommutative supersymmetric  Yang-Mills theory by
the ordinary Dirac-Born-Infeld theory with a magnetic field 
background\cite{seiberg,shiappa,cornalba,ishibashi,garousi}. Understanding this relation
 was originally 
one of the main purpose of this note, but no progress 
is made in this direction. 

Our analysis is expected to be 
directly generalized to the cases of  dyons and 1/4 BPS dyons. 
In addition, the leading effect of noncommutativity to the  moduli dynamics
of monopoles and 1/4 BPS dyons
will be of interest.

\centerline{\bf Acknowledgments}

%We acknowledge a useful discussion with S. Rey.  
This work 
is supported
in part by Ministry of Education Grant 98-015-D00061
and  KOSEF 1998 Interdisciplinary Research Grant
98-07-02-07-01-5.

\end{document}